\documentclass[prl,aps,twocolumn,showpacs,floatfix,reprint]{revtex4-1}
\usepackage{graphicx}
\usepackage{bm}
\usepackage{amsmath}
\usepackage{amssymb}
\usepackage{amsfonts}
\usepackage{dcolumn}
\usepackage{bm}
\usepackage{color}

\begin{document}

\title{Emergence of cooperativity in plasticity of soft glassy materials}

\author{Antoine Le Bouil}
\author{Axelle Amon}
\author{Sean McNamara}
\author{J\'er\^ome Crassous}

\affiliation{Universit\'e de Rennes 1, Institut de Physique de Rennes
  (UMR UR1-CNRS 6251), B\^{a}t.~11A, Campus de Beaulieu, F-35042 Rennes, France}

\date{\today}
\begin{abstract}
The elastic coupling between plastic events is generally invoked to
interpret plastic properties and failure of amorphous soft glassy
materials. We report an experiment where the emergence of a
self-organized plastic flow is observed well before the failure. For
this we impose an homogeneous stress on a granular material, and
measure local deformations for very small strain increments using a
light scattering setup. We observe a non-homogeneous strain that
appears as transient bands of mesoscopic size and well
defined orientation, different from the angle of the macroscopic
frictional shear band that appears at failure. The presence and the
orientation of those micro-bands may be understood by considering how
localized plastic reorganizations redistribute stresses in a
surrounding continuous elastic medium. We characterize the lengthscale
and persistence of the structure. The presence of plastic events and
the mesostructure of the plastic flow are compared to numerical
simulations.
\end{abstract}
\date{\today}
\maketitle

%%%%%%%%%%%%%%%%%%%%%%%%%%%%%%%%%%%%%%%%%%%%%%%%%%%%%%%%%%%%%%%%%%%%%%%%%%%%%%%%%%%%%%%%%%%%%%%%%%%%%%%%%
Amorphous materials have intermediate mechanical properties between
solids and liquids.  At low stress, they behave as elastic solids, but
deform plastically and flow when the stress increases.  These generic
behaviors, observed in many different systems such as concentrated
emulsions~\cite{Goyon2008}, colloidal systems~\cite{Besseling2010},
foams~\cite{Kabla2007b} or molecular glasses~\cite{Tanguy2006} with
apparently universal plastic or rheological
laws~\cite{Sollich1997,Derec2001}, suggest that such materials may be
described using a common
framework~\cite{Goyon2008,Katgert2010,Nguyen2011}. At the center of
those descriptions is the hypothesis of localized
reorganizations. Such events have been observed in many different
studies~\cite{Tanguy2006,Kabla2007b,Schall2007,Amon2012}. Each event
modifies locally the mechanical equilibrium, causing the surrounding
material to deform, and creating internal stresses. These stresses may
then provoke other events, leading to a succession or avalanche of
events~\cite{Falk1998,Maloney2006}. The coupling between events, and
its relevance to an avalanche-like cascade scenario for the
description of the final persistent shear-band is still an open
question~\cite{Maloney2006,Dahmen2011}.

Several experimental works show isolated reorganizations followed by localized flow structures, suggesting the existence of such coupling. Conclusions remain elusive in direct
observation of colloidal glasses due to the dominance of thermal
activity over the triggered events~\cite{Schall2007}.  In athermal
systems such as granular materials~\cite{Amon2012} or
foams~\cite{Kabla2007b}, the steps between accumulation of individual
events and appearance of shear bands remain unclear. Very recent
numerical and theoretical results suggest that reorganization
events may indeed couple in order to produce
bands~\cite{Maloney2006,Tsamados2008,Talamani2011,Martens2012,Dasgupta2012,Gimbert}.
However, the bands observed numerically resulting from the interacting
local events are transient and correspond to self-healing
micro-cracks, of a different nature than the final persistent
shear-bands. To our knowledge such transient micro-bands forming a
clear intermittent structure have never been reported experimentally.

We present in this letter the first direct experimental evidence
showing the progressive emergence of cooperative effects during
plastic deformations of an amorphous material. For this, we use a very
sensitive light scattering setup to monitor the homogeneous biaxial
compression of a granular material. We then show that the plastic flow
at the early stage of the loading of a granular material is concentrated
along self-healing micro-bands. The orientation of those transient
micro-bands are clearly different from the Mohr-Coulomb angle of the
final permanent shear band. We show that the orientations of those
microbands are given by the Eshelby solution~\cite{Eshelby1957} for the
long-range stress redistribution induced by local plastic
reorganizations in an elastic material.  We also show that the transient micro-bands are more  prominent  as the rupture is
approached.

%%%%%%%%%%%%%%%%%%%%%%%%%%%%%%%%%%%%%%%%%%%%%%%%%%%%%%%%%%%%%%%%%%%%%%%%%%%%%%%%%%%%%%%%%%%%%%%%%%%%%%%%%%%%%%%%%%%%%%%%%%%%%
\begin{figure}[tbh]
\includegraphics[width=\columnwidth]{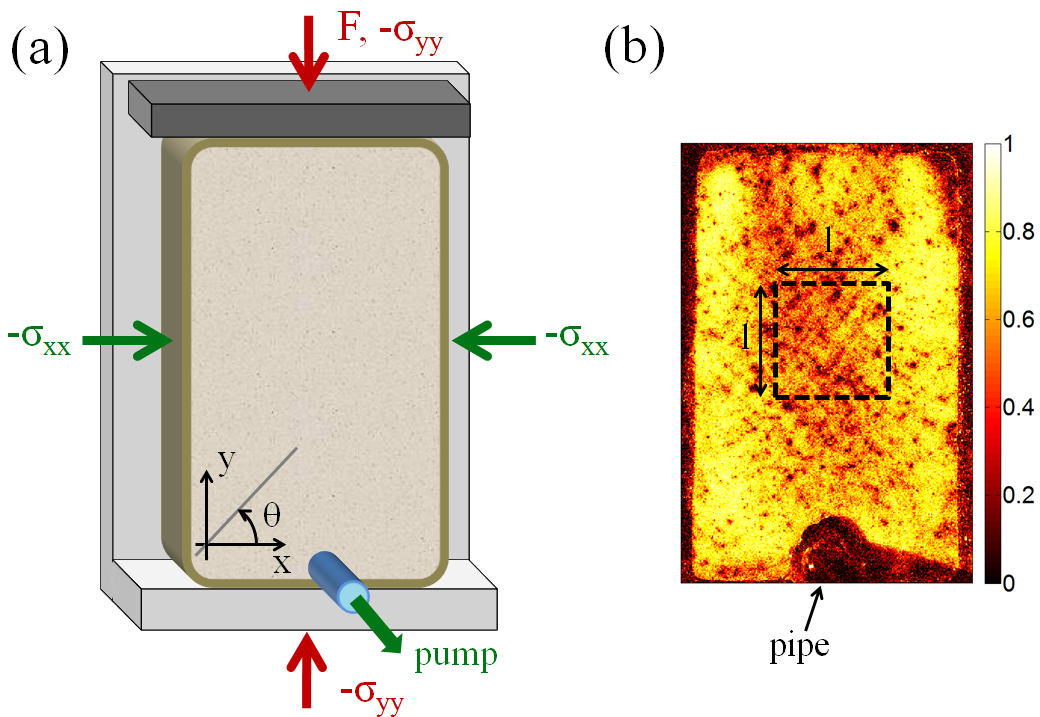}
\caption{(a) Schematic representation of the biaxial setup. The
  granular material is enclosed between a latex membrane and a glass
  plate (not represented here). A partial vacuum inside the membrane
  creates a confining stress $-\sigma_{xx}$. The sample is compressed
  at fixed velocity along the $y$ axis through a moving plate (upper
  plate, dark grey). The light grey back plate as well as the glass
  plate at the front forbid displacements along the $z$ direction
  ensuring plane-strain conditions. For compression,
  $-\sigma_{xx},-\sigma_{yy}>0$. (b) A map of correlation
  $g_I(\epsilon, {\bf r})$ with colorscale. The dashed area of side
  $l\simeq 270 d$ is the region of interest for the spatial correlation
  calculation.}
\label{fig1}
\end{figure}

\begin{figure}[tbh]
\includegraphics[width=\columnwidth]{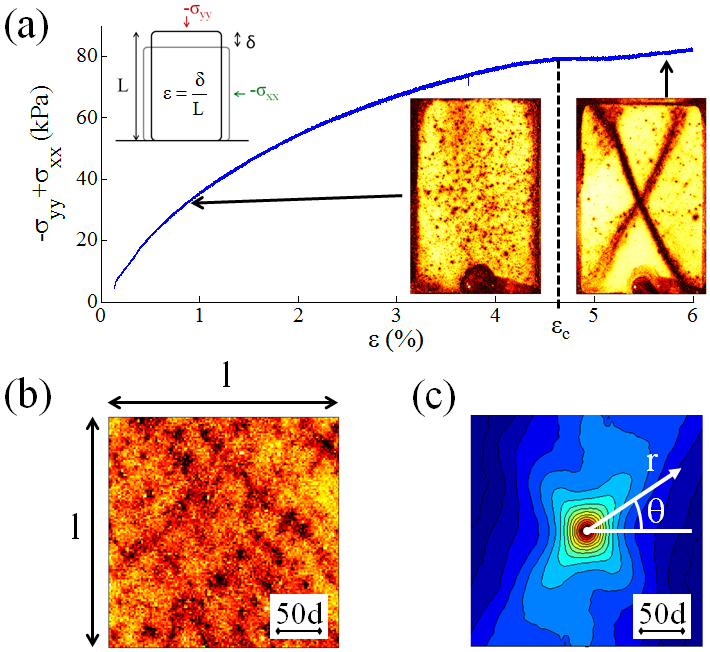}
\caption{(a) Applied stresses difference versus deformation
  ($-\sigma_{xx}=30$~kPa). Insets: left, notations; right: maps of
  $g_I(\epsilon, {\bf r})$ before failure
  ($\epsilon=-\epsilon_{yy}=0.91\%$) and after failure
  ($\epsilon=5.82~\%$). (b) Zoom of the region of interest of the
  deformation map before failure ($\epsilon=3.30~\%$) showing the mesoscale strain heterogeneities. (c) Correlation function $\Psi^{(0)}(\epsilon,\bf{r})$
  of $g_I$ at $\epsilon=3.30\%$ showing the plastic flow structure
  in a square of size $l \simeq 270d$ in the $\mathbf{r}$ plane.}
\label{fig2}
\end{figure}

--{\it Experimental setup.}  We deform an assembly of glass spheres by
imposing a homogeneous stress with a biaxial apparatus.  We recall
here the main features of the setup described extensively
elsewhere~\cite{LeBouil2013}: The material (glass beads, diameter
$d=90 \pm 20\mu m$, volume fraction $\approx 0.60$) is placed between
a preformed latex membrane (size $85 \times 55 \times 25$~mm) and a
glass plate. A pump produces a partial vacuum inside the membrane,
creating a confining stress $-\sigma_{xx}$.  The confined sample is
positioned on a metallic structure (in light grey on
Fig.~\ref{fig1}(a)). The glass plate is not represented on
Fig.~\ref{fig1}(a) and is at the front. The back metallic plate and
the front glass plate forbid displacement normal to the $xy$ plane,
ensuring plane-strain conditions. The bottom of the sample rests on a
fixed plate, while the upper plate (dark grey) is displaced by a step
motor.  The stress on the moving plate is
$-\sigma_{yy}=-\sigma_{xx}+F/S$, where $F$ is the force measured by a
sensor fixed to the plate, and $S$ the section of the sample. Although
there is probably some solid friction between the granular material
and the plates, we do not observe noticeable differences of
deformation between the upper and lower part of the sample.  The
stress gradient due to gravity is negligible, and the value of confining stress is such that cohesion effects and crushing of particles are unimportant. The global macroscopic
deformation is calculated as $\epsilon=-\epsilon_{yy}=\delta/ L$ with
$\delta$ the upper plate displacement and $L$ the sample height (see
left inset of Fig.~\ref{fig2}(a)). The compressions are done at fixed
deformation rate $d \epsilon / dt=1.1 \times 10^{-5}s^{-1}$. We
checked that we were in the quasistatic limit.

Strain heterogeneities are observed using a dynamic light scattering
setup~\cite{Erpelding2008}. An expanded $532$~nm laser beam
illuminates the material. Because of the coherence of the light
source, interferences occur and a speckle pattern forms. The image of
the front side of the sample is recorded by a $7360 \times 4912$
camera. Two different speckle images are compared using a correlation
method explained elsewhere~\cite{Erpelding2008}. Images are subdivided
in square zones, and for each zone we calculate the normalized
correlation function
\begin{equation}
g_I^{(1,2)}={\langle I_1I_2 \rangle - \langle I_1 \rangle \langle I_2 \rangle
\over \sqrt{ \langle I_1^2\rangle - \langle I_1 \rangle ^2}
	\sqrt{\langle I_2^2\rangle- \langle I_2\rangle^2}}
\end{equation}
where $I_1$ and $I_2$ are the intensity matrices of a same zone in two
different images, and $\langle\ldots\rangle$ indicates the average
over the zone.  Each zone becomes a pixel in a correlation map (see
Fig.~\ref{fig1}(b) and movie in supplemental material~\cite{movie}),
corresponding to a volume of surface $\sim 2.1d \times 2.1d$ in the $x-y$
plane and of depth of few $d$.  The decorrelation of the scattered
light comes from relative bead motions. We thus measure a combination
of affine and nonaffine bead displacements, and rotation of
non-spherical beads. In the
following we present maps based on images made at sample deformations
$\epsilon$ and $\epsilon+3.2 \times 10^{-5}$, and we note
$g_I(\epsilon, \mathbf{r})$ the value of the normalized correlation at
compression $\epsilon$ and at position $\mathbf{r}$ (see
Fig.~\ref{fig1}(b)).

%%%%%%%%%%%%%%%%%%%%%%%%%%%%%%%%%%%%%%%%%%%%%%%%%%%%%%%%%%%%%%%%%%%%%%%%%%%%%%%%%%%%%%%%%%%%%%%%%%%%%%%%%%%%%%%%%%%%%%%%%%%%%

--{\it Plastic flow structure.} Fig.~\ref{fig2}(a) shows the evolution
of the stress difference $\sigma_{xx}-\sigma_{yy}$ as a function of
the deformation $\epsilon$. At the beginning of the loading,
$\sigma_{xx}-\sigma_{yy}$ increases with $\epsilon$, and then attains
a plateau, consistent with numerous preceding studies, where a
granular material was prepared near the critical state volume
fraction~\cite{schofield}.  The stress plateau at $\epsilon_c=4.66 \%$ corresponds to the
failure of the sample, confirmed by the correlation map shown in
Fig.~\ref{fig2}(a) (rightmost inset). The deformation is dominated by
two symmetric shear bands where $g_I(\epsilon,{\bf r})$ is low,
corresponding to highly localized deformation. The inclination of the
bands is $\theta\simeq 65^{\circ}$, in agreement with a Mohr-Coulomb analysis
$\theta_{MC}=45+\varphi_c/2 \simeq 63^{\circ}$ for a frictional material,
 with $\varphi_c$ the internal friction angle~\cite{nedderman}. $\varphi_c =\arcsin[(\sigma_{yy}-\sigma_{xx})/(\sigma_{yy}+\sigma_{xx})]$ at failure ($\epsilon=\epsilon_c$). Those bands are permanent in the sense that they do not evolve with $\epsilon$ once they appear
(see movie in supplemental material~\cite{movie}.

Fig.~\ref{fig2}(b) shows a map of deformation before
failure. The deformation is strongly heterogeneous with a complicated
fine structure at small scale.  In contrast with
the permanent shear bands observed after failure, this deformation
pattern fluctuates strongly during the loading (see movie in
supplemental material~\cite{movie}).  To investigate the spatial
structure and intermittency of the plastic flow, we consider the
spatial correlation function of $g'_I \equiv 1 - g_I$:
\begin{align}
\Psi^{(\Delta \epsilon)}(\epsilon,&\mathbf{r})=\left\langle g_I'(\epsilon+\Delta \epsilon/2,\mathbf{r}')
      g_I'(\epsilon-\Delta \epsilon/2,\mathbf{r}+\mathbf{r}')\right\rangle\notag\\
	-&\left\langle g_I'(\epsilon+\Delta \epsilon/2,{\mathbf{r}'})\right\rangle \left\langle g_I'(\epsilon-\Delta \epsilon/2,{\mathbf{r+r}'})\right\rangle
\end{align}
where $\langle\ldots\rangle$ is an average over $100$ correlation
maps, i.e. a deformation of $3.2\times10^{-3}$, and over $\mathbf{r}'$, for $\mathbf{r}'$ and $\mathbf{r+r}'$
covering the region of interest on
Fig.~\ref{fig1}(b). Fig.~\ref{fig2}(c) shows a plot of $\Psi^{(0)}(\epsilon,\mathbf{r})$. Along two symmetric directions $\theta= \pm \theta_E$
with $\theta_E\approx 53^\circ$ the correlation decays slowly with
$r$ (see fig.4.(a)). The direction of anisotropy $\theta_E$ is almost constant during
the loading, and is clearly different from $\theta_{MC}$.

%%%%%%%%%%%%%%%%%%%%%%%%%%%%%%%%%%%%%%%%%%%%%%%%%%%%%%%%%%%%%%%%%%%%%%%%%%%%%%%%%%%%%%%%%%%%%%%%%%%%%%%%%%%%%%%%%%%%%%%%%%%%%
\begin{figure}[tbh]
\includegraphics[width=\columnwidth]{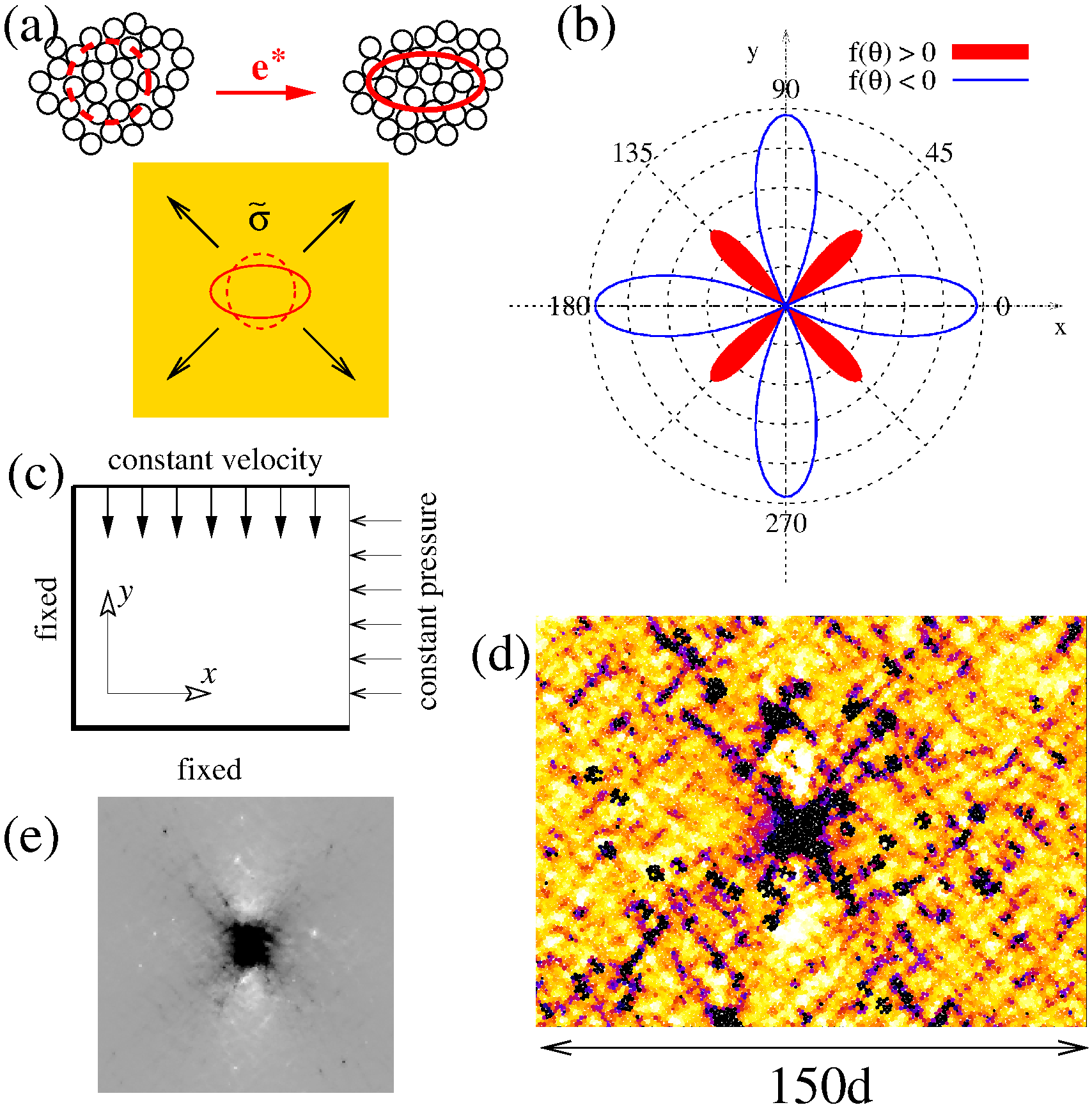}
\caption{(a) Schematic representation of a local plastic event
  specifying the tensors $\mathbf{e^*}$ (linked to the
  deformation of the inclusion) and $\tilde{\bm{\sigma}}$ (stress
  redistribution in the surrounding medium due to the plastic
  event). (b) Angular distribution of
  $\tilde\sigma_{xx}-\tilde{\sigma}_{yy}\propto f(\theta)$ in the case
  of an isovolumic transformation of the inclusion ($\nu = 0.33$). (c)
  Boundary conditions of the numerical simulations. (d) Example of a
  deformation map from numerical simulation displaying a local event and micro-bands. (e) Synthetic local reorganization obtained numerically
  by a modification of the elastic constants of few grains.}
\label{fig3}
\end{figure}

--{\it Localized plastic events.} To explain the observed structure of
the plastic flow we first investigate theoretically the consequences of a single,
isolated reorganization somewhere in the granular material.  Consider
a plastic deformation that relaxes stress within a small volume, but
redistributes it in the surrounding material. We consider that the
surrounding region behaves as a linear elastic
material~\cite{Makse2004}, that we will suppose isotropic with Poisson
ratio $\nu$.  Eshelby gave an analytical solution to this 3D
problem~\cite{Eshelby1957}: Let $\mathbf{e^*}$ be the strain tensor
of the reorganization (see Fig.\ref{fig3}(a)).
 We suppose $e^*_{xy}=0$, i.e., that
$\mathbf{e^*}$ is coaxial to the applied stress tensor and
$e^*_{zz}=e^*_{xz}=e^*_{yz}=0$ because of the plane-strain
configuration, leaving only $e^*_{xx}$ and $e^*_{yy}$ as the
non-zero strain components.
Far from the rearrangement, the additional stress
originating from the rearrangement in the x-y plane is $\tilde{\bm{\sigma}}$, with
$\tilde{\sigma}_{xx}-\tilde{\sigma}_{yy}\propto f(\theta)$, where
\begin{align}
f(\theta) =& (e^*_{xx}-e^*_{yy}) \left[-\frac{15}{4} \cos(4\theta) +
  \frac{8\nu - 7}{4} \right] \notag \\ & - \frac{9}{2} (e^*_{xx} +
e^*_{yy}) \cos(2\theta).
\label{eq:Eshelby}
\end{align}
If $\tilde{\sigma}_{xx}-\tilde{\sigma}_{yy}>0$ the redistributed
stress adds to the applied stress, increasing strain along those
directions. Its maximum occurs for
$\cos(2\theta_E^*)=\frac{3}{10}\frac{e^*_{yy} + e^*_{xx}}{e^*_{yy} -
  e^*_{xx}}$. In the case of an isovolumic transformation,
$\theta^*_E=45^\circ \pmod {90^\circ}$. Fig.~\ref{fig3}(b) shows $f(\theta)$
in this case. For a local rearrangement in agreement with the
macroscopic deformation of the sample, i.e. $e^*_{xx}$ and $e^*_{yy}$
of opposite signs, $\theta_E^*$ increases (resp. decreases) for a
dilating (resp. contracting) rearrangement, with extremal values
$\frac{1}{2}\cos^{-1}(\pm 3/10)$. The largest possible value for
$\theta_E^*$ is then $54^\circ$, close to the value of $\theta_E\approx 53^\circ$ of the experiment.  This reorganization structure
has been shown in numerical studies of molecular
glasses~\cite{Maloney2006,Tsamados2008} and cellular
foam~\cite{Kabla2007b}, but the existence of such elastic
redistribution in frictional granular material is still an open
question. Indeed, the existence of an elastic limit for such system is
still a matter of debate~\cite{Makse2004}. We performed numerical
bidimensional Discrete Element Method simulations of a biaxial compression test
(see Fig.~\ref{fig3}(c) for boundary conditions). Fig.~\ref{fig3}(d)
shows results from a simulation of $N=256^2$ grains, using a
visualization method inspired by the experimental technique: Positions
of the grains are recorded at strain increments of $\delta \epsilon =
10^{-5}$. Two successive system states are compared, and for each
grain, a local strain (average relative change in distance to its
neighbors) is calculated.  Those grains whose local strain is large
are dark. We can generate a plastic event in the simulation by
softening a small number of grains in the sample (see
Fig.~\ref{fig3}(e)) and we obtain a local deformation in accordance
with the analytical solution of Fig.~\ref{fig3}(b). Fig.~\ref{fig3}(d) shows that such local events also occur
during the compression of the granular material.

--{\it Coupling between localized events and plastic flow structure.}
Along the
directions where $\tilde{\sigma}_{xx}-\tilde{\sigma}_{yy}$ is positive,
the additional stress has the same sign as the applied stress,
possibly triggering new reorganizations. We therefore expect
deformation to be organized in micro-bands whose orientations are given by
the Eshelby solution. This structure is visible in the
numerical experiments where very transient localized lines inclined at
$\theta \approx \pm 45^\circ$ are present (see Fig.~\ref{fig3}(d)). The resulting
images display the same phenomenology as the experimental results:
well before failure, deformation is concentrated in short diagonal
micro-bands, (probably similar to those reported in other studies~\cite{Kuhn1999,Gimbert,Hall2010}), and at failure, a shear band appears (not shown
here). The agreement between 2D simulation and 3D experiments supports our plane strain hypothesis.

\begin{figure}[tbh]
\includegraphics[width=\columnwidth]{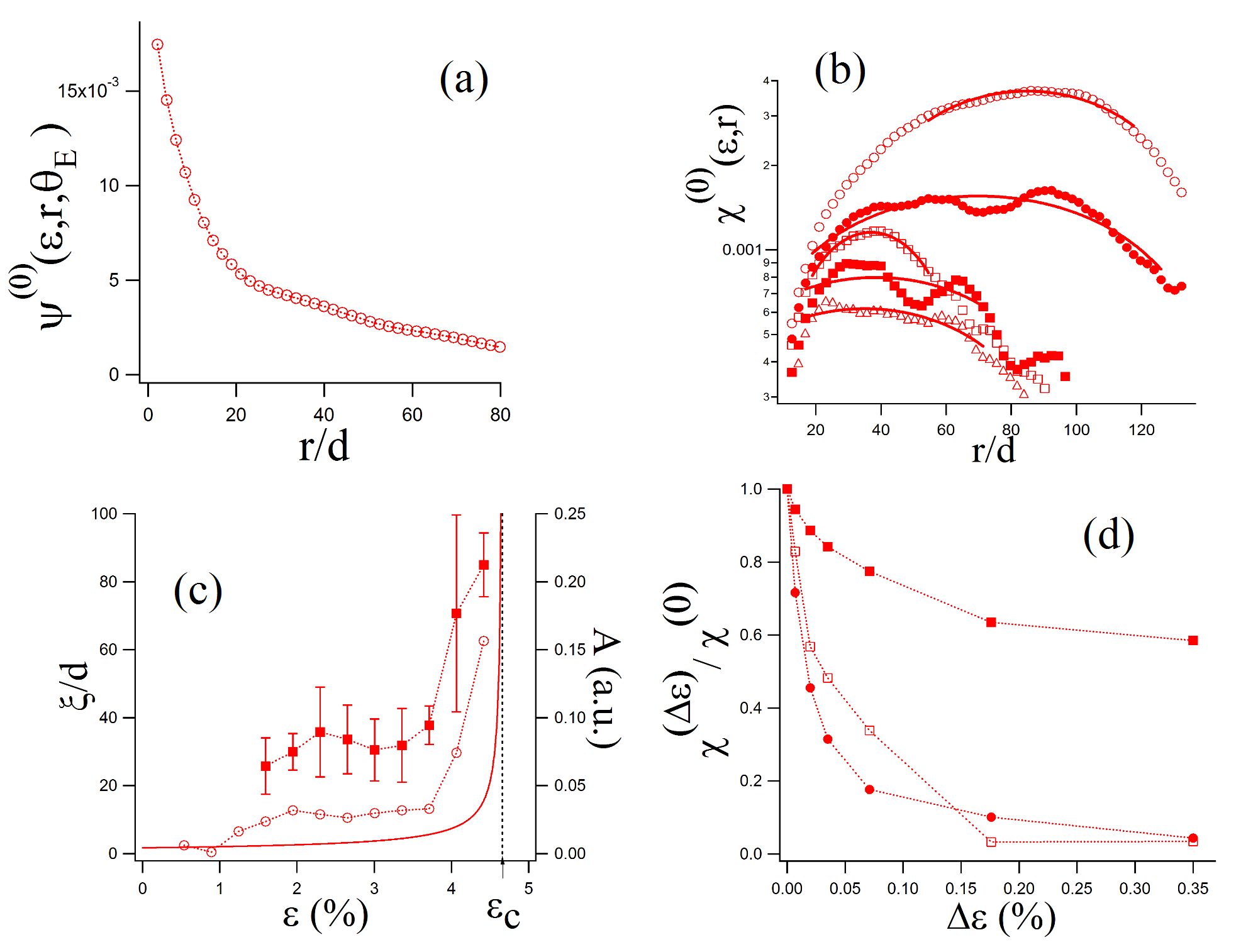}
\caption{(a) $\Psi^{(0)}$ as function of $r$ for $\epsilon=2.7\%$, showing a fast decay at short distance ($r/d \lesssim 10$) followed by a slow decay. (b) $\chi^{(0)}(\epsilon,r)$ versus $r/d$ for increasing values of deformations $\epsilon=1.6\%~(\triangle)$,
  $\epsilon=2.3\%~(\blacksquare)$, $\epsilon=3.7\%~(\square)$,
  $\epsilon=4.0\%~(\bullet)$, $\epsilon=4.4\%~(\circ)$. Lines are quadratic fit around maximum. (c) length
  $\xi/d~(\blacksquare)$ and mean amplitude $A~(\circ)$ (see text) as functions
  of the deformation $\epsilon$. Error bars are given by uncertainly of the quadratic fit of $\chi^{(0)}$ around maximum.  The black dotted line indicates the
  deformation at rupture $\epsilon_c\approx4.66\%$. Plain line is the cooperative length~\cite{notemu} expected from non-local flow rule of granular material~\cite{kamrin2013}. (d) Relaxation of
  $\bigl(\chi^{(\Delta \epsilon)}/\chi^{(0)}\bigr)(\epsilon,\xi(\epsilon))$ as a function of the deformation
  increment $\Delta \epsilon$ for
  $\xi(\epsilon=3.3\%)=33d~(\bullet)$,
  $\xi(\epsilon=4.0\%)=70d~(\square)$ and
  $\xi(\epsilon=4.4\%)=85d~(\blacksquare)$.}
\label{fig4}
\end{figure}

--{\it Spatial and temporal correlations.}
Coming back to our experimental data, we focus on the evolution of the
anisotropic part of $\Psi^{(\Delta
  \epsilon)}(\epsilon,r,\theta)=\Psi^{(\Delta
  \epsilon)}(\epsilon,\mathbf{r})$ during the loading, which we define
as:
\begin{align}
\chi^{(\Delta \epsilon)}(\epsilon,r)=&{1 \over 2}[\Psi^{(\Delta
    \epsilon)}(\epsilon,r,\theta_E)+\Psi^{(\Delta
    \epsilon)}(\epsilon,r,-\theta_E)]\notag\\ -&\Psi^{(\Delta
  \epsilon)}_{iso}(\epsilon,r)
\end{align}
with $\Psi^{(\Delta \epsilon)}_{iso}(\epsilon,r)={1\over 2
  \pi}\int_0^{2 \pi} \Psi^{(\Delta
  \epsilon)}(\epsilon,r,\theta)d\theta$ the isotropic part of
$\Psi^{(\Delta \epsilon)}$. Fig.~\ref{fig4}(b) shows the evolution of
$\chi^{(0)}(\epsilon,r)$ in function of $r$ for different values of
$\epsilon$. We observe that the anisotropic part of the correlation
function increases as the loading increases. We consider a two-fold
characterization of $\chi^{(0)}$. First the integral
$A(\epsilon)=\int_{r=0}^{r=\ell/2}\chi^{(0)}(\epsilon,r)
dr$ estimates the strength of the anisotropy.
Second the characteristic distance $\xi(\epsilon)$
at which the correlation is maximum $\bigl({\partial \chi^{(0)}/\partial r}\bigr)\bigl(\epsilon,\xi(\epsilon)\bigr)=0$ is computed using a quadratic
fit (plain line of fig.~4(b)) of the experimental curves near maximum. Fig.~\ref{fig4}(c) shows that both the
integral $A$ and the characteristic length $\xi/d$
of the anisotropy increase as
the loading progresses toward rupture. Finally, the transient nature
of the observed structure can be shown by considering the scale of
deformation at which the plastic flow persists. For this, we
considered the evolution of $\bigl(\chi^{(\Delta
  \epsilon)}/\chi^{(0)}\bigr)(\epsilon,\xi(\epsilon))$ with $\Delta
\epsilon$ at a given $\epsilon$. Fig.~\ref{fig4}(d) shows that close
to rupture, for $\epsilon=4.4\%$ and $\xi(\epsilon=4.4\%)=85d$, the
deformation persists after a deformation increment $\Delta
\epsilon\approx 0.3\%$. On the contrary, further from the failure
($\epsilon=3.3\%$), the deformation
decays over a typical increment of deformation $\Delta \epsilon\approx
0.02\%$.

From the structure of the plastic flow, a characteristic length $\xi$ revealing the cooperativity of the fluctuation of plastic flow emerges.
The values of $\xi$ are in quantitative agreement with numerical simulations of granular material~\cite{Kuhn1999} where fluctuations coupled on distance $\sim 10-40d$ are reported. Theoretically~\cite{bocquet2009} a non-local rule for the mean plastic flow is expected to emerge from those fluctuations. Such non-local flow rule have been proposed to describe granular plastic flow~\cite{Kamrin2012}. Fig.~4(c) shows the expected evolution of the cooperativity length~\cite{notemu} proposed in~\cite{kamrin2013} during the loading. The cooperativity length of the mean flow is smaller than $\xi$. This is probably due to the coarse-graining process described in~\cite{bocquet2009}.

%%%%%%%%%%%%%%%%%%%%%%%%%%%%%%%%%%%%%%%%%%%%%%%%%%%%%%%%%%%%%%%%%%%%%%%%%%%%%%%%%%%%%%%%%%%%%%%%%%%%%%%%%%%%%%%%%%%%%%%%%%%%%
--{\it Conclusion.} In summary, a careful experimental study of the
plastic flow of an athermal amorphous material reveals a mesoscopic
structure of the strain since the early stage of the loading process:
deformation concentrates in transient short micro-bands of well-defined
orientation. We connect those orientations with the
elastic long-range stress redistribution due to localized plastic
reorganizations. We show an
increasing characteristic length and persistence during
the loading. However, the relationship between
these transient micro-bands and the final permanent frictional
shear bands is more complex than the description of a final persistent
shear-band formation as a mere growing cascade of local
rearrangements.
The final shear band does not arise from a coalescence of micro-bands,
nor is it initiated by a single micro-band that reaches the boundary
and becomes locked.
Instead, as the movie in supplemental material~\cite{movie} shows,
the two types
of deformation, oriented in two different directions, coexist near
failure. We observe a hierarchical structure with a mesoscopic pattern embedded in large scale shear band. The modelization of the final persistent shear-band
needs to describe the complex interaction between
the micro-bands and the larger scale localization. The
careful characterization of the birth of the permanent shear band is a
work-in-progress.

%%%%%%%%%%%%%%%%%%%%%%%%%%%%%%%%%%%%%%%%%%%%%%%%%%%%%%%%%%%%%%%%%%%%%%%%%%%%%%%%%%%%%%%%	
This work has been supported by ANR (No.2010-BLAN-0927-01) and
R\'egion Bretagne (MideMade). We thanks P.~Chasle, H.~Orain,
J.-C.~Sangleboeuf, P.~B\'esuelle and C.~Viggiani for help with the
biaxial apparatus, and GDR Mephy for fruitful discussions.

%%%%%%%%%%%%%%%%%%%%%%%%%%%%%%%%%%%%%%%%%%%%%%%%%%%%%%%%%%%%%%%%%%%%%%%%%%%%%%%%%%%%%%%%

\end{document}